\documentstyle[12pt]{article}
\def\a{\alpha}

\def\g{\gamma}
\def\d{\delta}
\def\e{\epsilon}
\def\lam{\lambda}
\def\ro{\rho}
\def\si{\sigma}
\def\H{{\rm\scriptscriptstyle H}}
\def\T{{\rm\scriptscriptstyle T}}
\def\pt{p_\T}
\def\ph{p_\H}
\def\mh{M_\H}
\def\frac#1#2{{#1 \over #2}}
\def\slash#1{\rlap/#1}
\def\GeV{{\rm GeV}}
\def\TeV{{\rm TeV}}
\def\la{\langle}
\def\ra{\rangle}
\def\l{\left}
\def\r{\right}
\def\M{{\cal M}}
\def\o{\over}
\def\beq{\begin{equation}}
\def\beqa{\begin{eqnarray}}
\def\eeq{\end{equation}}
\def\eeqa{\end{eqnarray}}
\def\ch{\leftrightarrow}
\def\pe{\stackrel{P}{=}}
\def\ce{\stackrel{C}{=}}
\def\be{\stackrel{B}{=}}
\def\cpe{\stackrel{CP}{=}}
\def\bpe{\stackrel{BP}{=}}
\def\plim{\stackrel{\ph\to0}{\to}}
\def\tr{{\rm tr}}
\def\no{\nonumber}
\def\noi{\noindent}
\def\qq{\qquad}
\def\re{{\rm Re}}
\begin{document}
\begin{tabbing}
~~~~~~~~~~~~~~~~~~~~~~~~~~~~~~~~~~~~~~~~~~~~~~~~~~~~~~~~~~~~~~~~~~~~~~
\= F\&M96/3 \\
\> hep-ph/9610541 
\end{tabbing}
\vskip 1cm
\begin{center}
{\bf PRODUCTION OF A HIGGS BOSON PLUS TWO JETS IN HADRONIC COLLISIONS} \\
\vskip 1cm
{\bf  Russel P. Kauffman,~ Satish V. Desai and Dipesh Risal} \\
\vskip 1.cm
{\it Department of Physics and Astronomy} \\
{\it Franklin and Marshall College, Lancaster, PA~~17604} \\
\vskip 1cm
ABSTRACT
\end{center}
\vskip 0.5cm
\noindent
We consider the production of a Standard Model Higgs boson accompanied
by two jets in hadronic collisions.  
We work in the limit that the top quark is much heavier
than the Higgs boson and use an effective Lagrangian for the 
interactions of gluons with the Higgs boson.  In addition to the 
previously computed four-gluon process, we compute the the amplitudes
involving two quarks, two gluons and the Higgs boson and those involving
four quarks and the Higgs boson.  
We exhibit the form of our results in the small-$\ph$ and factorization
limits.  We present numerical results for 
$\sqrt{S}= 14~\TeV$ and $\sqrt{S}= 2~\TeV$.  
We find that the dominant processes are $gg \to ggH$ and $qg \to qg H$ with
the former (latter) contributing about 60\% (40\%) of the cross section at 
$\sqrt{S}= 14~\TeV$ and the two processes each contributing about half the 
cross section at $\sqrt{S}= 2~\TeV$.  All other processes are negligible at 
both energies.
\vskip0.25cm
\noindent September 1996\hfill
\vfill
\pagebreak

\section{Introduction}

The Higgs boson is the last remaining undiscovered element of the   
Standard Model.  Discovery of a Higgs boson (or more than one)
would confirm that the
Higgs mechanism is the source of electro-weak symmetry breaking while
convincing evidence that Higgs bosons do not exist would necessitate
another explanation for electro-weak symmetry breaking. 
Thus, the search for the Higgs boson is one of the fundamental quests 
of modern
high energy physics.  Current published
experimental results set a lower limit on the Higgs boson mass of about
60 GeV
\cite{mhlimit}
while the $e^+e^-$ collider  LEPII can be expected to extend this limit
to somewhere near $80~\GeV$.

  In this paper, we are concerned with 
the production of Higgs bosons in hadronic collisions.
We are particularly interested in the 
 so-called ``intermediate mass'' Higgs boson, {\it i.e.}, one in
the mass region $80~\GeV \leq\mh \leq 200~\GeV$,
although, as we will argue, our results have a somewhat larger range
of applicability.   Experimentally
this is an extremely difficult region in which to see the Higgs boson, due
to the large backgrounds to the common Higgs
boson decay channels.  
Hence it is vital to have precise predictions for the
production cross section as well as for the distribution of the produced
Higgs bosons in transverse momentum and rapidity.  The probability
of extra particles, be they jets, W's or top quarks,
being produced along with the Higgs boson,
also impacts its detection.  Accompanying particles 
may act as tags or be confused with the Higgs decay products.

Here  we  discuss the production of the Higgs boson accompanied by 
two jets.  The cross section for $g g \to ggH$ was calculated previously 
\cite{hgggg}.
We compute the contributions needed for the total Higgs boson plus two jet
cross section: $gg\to qqH$, $qg \to qgH$, $qq \to ggH$, $qq \to qqH$,
where `$q$' stands generically for a quark or anti-quark of undetermined
flavor. We consider only the QCD generated processes, that is, extra
quark or gluons lines attached to the basic gluon-gluon--Higgs-boson
interaction. (The electroweak process, $q q \to qq H$ through $WW$ or $ZZ$
fusion does not interfere with the all quark process considered here. The 
interference term would be proportional to the trace of a single $SU(3)$ 
generator and so vanishes.)

We will work in the limit in which the top quark is much heavier
than the Higgs boson and all of the energy scales in the problem.  
Since experiments \cite{cdf}
place $m_{\rm top}\simeq 175~\GeV$, 
this limit is relevant for the intermediate mass Higgs boson.  
This limit is also
relevant for consideration of soft and collinear radiation surrounding
the production of the Higgs boson.

The organization of the paper is as follows.  The effective Lagrangian is
discussed in Section 2.  Section 3 contains the spinor product formalism
in which the amplitudes will be computed.  The amplitudes for a Higgs boson
plus two or three massless particles are computed in Section 4.  For 
completeness the Higgs boson plus four gluon amplitude is presented in 
Section 5.  Sections 6 and 7 contain the calculations of the amplitude 
involving a Higgs boson
plus a quark anti-quark pair and two gluons and the amplitude for a 
Higgs boson plus two quark anti-quark pairs, respectively.  The limit
of our results when the momentum of the Higgs is small is presented in 
Section 8 and their behaviour in the factorization limits is presented
in Section 9. Section 10 contains our numerical results and the Appendix
contains the squares of the various amplitudes.

\section{The Effective Lagrangian}

The production mechanism in which we are interested is
$gg \rightarrow H$ which occurs
through a quark loop where   the only numerically
important contribution  
is that of the top
quark. In the limit in which
the top quark is heavy,
 $m_{\rm top} \gg \mh$, the cross section can be computed via the
following effective Lagrangian \cite{rusk}
\begin{equation} 
{\cal L}_{\rm eff}=-{1\over 4} A H G^A_{\mu \nu} G^{A~\mu \nu},
\label{eq:leff}
\end{equation}
where $G^A_{\mu \nu}$ is the field strength of the SU(3) color
gluon field and $H$ is the Higgs-boson field.  The effective coupling $A$
is given by $A = \a_s /(3 \pi v)$, where $v$ is the vacuum expectation
value parameter, $v^2=(G_F\sqrt{2})^{-1}=(246~\GeV)^2$.
The effective Lagrangian generates vertices involving the Higgs boson
and two, three or four gluons.  The associated Feynman rules are displayed
in Fig 1.  The two-gluon--Higgs-boson vertex is proportional to the
tensor 
\begin{equation} 
H^{\mu\nu}(p_1,p_2) = g^{\mu\nu}p_1 \cdot p_2 - p_1^\nu p_2^\mu.
\label{eq:bigh}
\end{equation}
The vertices involving three and four gluons and the Higgs boson 
are proportional to their counterparts from pure QCD:
\begin{equation} 
V^{\mu\nu\ro}(p_1,p_2,p_3) = (p_1-p_2)^\ro g^{\mu\nu}  
                             + (p_2-p_3)^\mu g^{\nu\ro}
                             + (p_3-p_1)^\nu g^{\ro\mu},    
\label{eq:bigv}
\end{equation}
and
\begin{eqnarray} 
X^{\mu\nu\ro\si}_{abcd} &=&  
f_{abe}f_{cde}( g^{\mu\ro}g^{\nu\si} - g^{\mu\si}g^{\nu\ro} )
+f_{ace}f_{bde}( g^{\mu\nu}g^{\ro\si} - g^{\mu\si}g^{\nu\ro} ) 
\nonumber \\ 
&+&f_{ade}f_{bce}( g^{\mu\nu}g^{\ro\si} - g^{\mu\ro}g^{\nu\si} ).
\label{eq:bigx}
\end{eqnarray}
It is straightforward to use this Lagrangian to obtain the ${\cal O}
(\alpha_s^3)$ contributions to the process $gg \rightarrow H$ 
\cite{sally,zerwas}.
These radiative corrections increase the 
lowest order rate by a factor of 1.5 to 2.  
As a by-product of the calculation of the ${\cal O}(\alpha_s^3)$
 radiative
corrections to $gg\rightarrow H$, one also 
obtains the cross section for $gg\rightarrow Hg $.

If the Higgs boson mass is of the same order as the top quark mass or larger
the approximation entailed in the effective Lagrangian breaks down.  However,
even if $\mh$ is not much smaller than $m_t$, 
the results of the effective
Lagrangian can be applied, after some modification, in the the soft
and/or collinear regime. 
Factorization requires that as a gluon becomes soft or two particles become
collinear an amplitude must factor into a divergent piece 
and a non-divergent piece, the divergent piece being
independent of the hard process. 
Applied to the case of the Higgs-jet-jet amplitudes,
when both outgoing jets have small $\pt$ (compared to the lowest scale in
the problem, $\mh$ or $m_t$) factorization requires that the dependence on 
$\mh/m_t$ must be the same as the lowest order $Hgg$ amplitude.  This was shown
explicitly in Ref. \cite{hpt} for the processes $gg\to gH$, $qg \to qH$ and 
$q\bar q \to gH$.  Since in this limit 
the only dependence on $m_t$ is in the overall
factor, the result derived from the effective Lagrangian may simply be rescaled
in order to be applied to the case when $m_t$ is not much larger than $\mh$.

\section{Spinor Product Formalism}
We are interested in processes in which all the particles except
the Higgs boson are massless.  Each amplitude can be expressed in
terms of spinors in a Weyl basis.  For light-like momentum $p$ and
helicity $\lam=\pm1$ we introduce spinors \cite{helic,ber}
\begin{eqnarray}
&&|p{\pm}\ra ={1\over 2} (1\pm \gamma_5)u(p) = 
{1\over 2} (1\mp \gamma_5)v(p)\nonumber \\
&&\la p{\pm}| = \overline{u}(p){1\over 2} (1\mp \gamma_5)
= \overline{v}(p){1\over 2} (1\pm \gamma_5).
\label{eq:spindef}
\end{eqnarray}
Polarization vectors for massless vector bosons can be written in
terms of these spinors.  For a gluon of 
momentum $k$ and positive or negative helicity
\begin{equation}
  \epsilon^{\mu}_{\pm} = { \la q{\pm}|\gamma^{\mu}|k{\pm}\ra
                 \over \sqrt{2}\la q{\mp}| k{\pm} \ra } \quad,
\label{eq:epsilons}
\end{equation}
where the reference momentum $q$ satisfies $q^2=0$ and $q\cdot k\neq 0$
but is otherwise arbitrary; the freedom of the choice of $q$ is a reflection
of gauge invariance.
Each helicity amplitude
can be expressed in terms of products of these spinors:
\begin{eqnarray}
\la p{-}|q{+} \ra &=& {-}\la q{-}|p{+} \ra \equiv \la pq \ra, \nonumber \\
\la p{+}|q{-} \ra &=& {-}\la q{+}|p{-} \ra \equiv [ pq ], \nonumber \\
\la p{-}|q{-} \ra &=& \la q{+}|p{+} \ra = \la pp \ra = [pp] = 0.
\label{eq:spinprod}
\end{eqnarray}
The spinor product $\la pq \ra$ and $[pq]$ are complex square roots
of $2\, p\cdot q$,
\begin{eqnarray}
 &&\la pq \ra [qp] =2\, p\cdot q, \nonumber \\
&&\la pq \ra^* = {\rm sign}(p\cdot q) [qp].
\label{eq:spconj}
\end{eqnarray}
Using the identities
\begin{equation}
\slash p = |p{+}\ra \la p{+} | + |p{-}\ra \la p{-} | 
\label{eq:helproj}
\end{equation}
and
\begin{equation}
\la p {\pm} | \g^\mu | q {\pm} \ra \g_\mu
= 2 (|q{\pm}\ra \la p{\pm} | + |p{\mp}\ra \la q{\mp} |, 
\label{eq:fierz}
\end{equation}
each amplitude can be written solely in terms of spinor products.
The following identity and its complex conjugate are
useful for simplifying the results:
\begin{equation}
\la pq \ra \la rs \ra = \la ps \ra \la rq \ra + \la pr \ra \la qs \ra.
\label{eq:rearrange}
\end{equation}

For the remainder of the paper we will use the convention that all the 
particles are outgoing.  The amplitudes for the
various processes involving two incoming 
massless particles and two outgoing massless particles plus a Higgs boson
can then be obtained by crossing symmetry. The momenta
of the massless particles are labelled $p_1,~p_2,~p_3,~p_4$ with the
Higgs boson momentum being $p_\H$.  Our convention is then
$p_1+p_2+p_3+p_4+\ph=0$. We will use the shorthand notations 
$\la p_i p_j \ra = \la ij \ra$, $ [p_i p_j ] = [ij]$, 
$(p_i + p_j)^2 = S_{ij}$, and $(p_i + p_j + p_k)^2 = S_{ijk}$.  

\section{Two and Three Particle Plus Higgs Boson Processes}

As a preliminary to calculating the Higgs plus four-parton amplitudes,
we present here the two and three parton amplitudes as examples.
These amplitudes also provide the limiting forms 
of the four-parton amplitudes when two partons become collinear or
a gluon becomes soft.  
The lowest order process by which Higgs bosons are created is
$g g \to H$.  The Feynman diagram is given in Fig. 1a.  The non-zero
helicity amplitudes are $++$ and $--$.  We find
\begin{equation}
\M^{++} = {1\o2} i A [12]^2 \d^{ab}.
\label{eq:hggamp}
\end{equation}
The amplitude for the $--$ helicity combination can be obtained by exchanging
square brackets for triangle brackets in this expression.

The lowest order processes which produce Higgs bosons with non-zero 
transverse momentum are $gg \to gH$, $qg \to qH$ and $q \bar q \to g H$.
The third process is a crossing of the second.  The relevant Feynmam 
diagrams are shown in Fig.'s 2 and 3. For $gggH$ 
the independent helicity amplitudes are
(labelled according to helicities of gluons 1, 2 and 3 in order)
\begin{eqnarray}
\M^{+++} &=& { g A f_{abc} \mh^4
 \o \sqrt{2} \la 12 \ra \la 23 \ra \la 31 \ra } \\
\M^{-++} &=& { g A f_{abc} [23]^3
 \o \sqrt{2} [12] [13]  }.
\label{eq:hgggamp}
\end{eqnarray}
The parity conjugate amplitudes can be obtained by exchanging square brackets
for triangle brackets and multiplying by -1.
Squaring these amplitudes and summing over helicities and colors
leads to the known result \cite{hinch,sally}
\begin{equation}
\sum |\M(H\to ggg)|^2 = { g^2 A^2 N(N^2-1) \o S_{12} S_{13} S_{23} }
\l( S_{12}^4 + S_{13}^4 + S_{23}^4 + \mh^8 \r),
\label{eq:hgggsum}
\end{equation}
where $N=3$ is the number of colors.

The helicity amplitudes for the process $H\to q\bar q g$ can be obtained
similarly.  Since QCD is helicity conserving in the massless limit, the 
quark and anti-quark must have opposite helicities.  Using the momentum
and color assignments in Fig. 3 we have (labeling the amplitudes by the
helicity of the quark, antiquark and gluon, in that order)
\beq
\M^{+-+} = -{ig T^a_{ij} A \o \sqrt{2} } {[13]^2 \o [12]},
\label{eq:hqqgamp}
\eeq
where the $SU(3)$ generators are
normalized such that $Tr(T^a T^b)={1\over 2}\delta_{ab}$.
To get the parity conjugate amplitude $\M^{-+-}$ exchange 
square brackets with triangle brackets.  To get the charge conjugates of these
two amplitudes, $\M^{-++}$ and $\M^{+--}$, respectively,
exchange $p_1 \leftrightarrow p_2$.  Again, when squared and summed over 
colors and helicities, these results agree with the known expression
\cite{hinch,sally}.

\section{The Higgs Boson Plus Four Gluon Amplitude}

The $Hgggg$ amplitude \cite{hgggg}
is obtained by summing the 26 
Feynamn diagrams detailed in Fig. 4.  Unlike the case of $Hggg$ not all
the diagrams have the same color structure.  To facilitate the 
cancellations that simplify the amplitude we introduce the {\it dual
color decomposition}.
The scattering amplitude for a Higgs boson and $n$ gluons with external
momenta $p_1$, ...$p_n$, colors $a_1$,...$a_n$, and
helicities $\lambda_1$,...$\lambda_n$ is written as 
\cite{parki,parkii,parkiii}
\beq
{\cal M}= 2Ag^{n-2}
\sum_{\rm perms} {\rm tr} (T^{a_1}...T^{a_n})m(p_1,\epsilon_1;
...;p_n,\epsilon_n),
\label{eq:dual}
\eeq
where the sum is over the non-cyclic permutions of the
momenta.  This form of the amplitude emerges when the identities 
\beqa
f_{abc} &=& -2i {\rm Tr}(T^a T^b T^c - T^c T^b T^a), \nonumber \\
f_{abe} f_{cde} &=& -2 {\rm Tr}([T^a,T^b][T^c,T^d])
\label{eq:fabc}
\eeqa
are used to replace the $f_{abc}$'s with traces of combinations of $T^a$'s.
The utility of the dual decomposition, Eq.~\ref{eq:dual}, 
comes from the properties
of the ordered sub-amplitudes $m(p_1,\epsilon_1;...;p_n,\epsilon_n)$, which we 
abbreviate $m(1,...,n)$:
\hfill\break
  \indent 1) they are invariant under cyclic permutations of the momenta;\hfill
\break
  \indent 2) they are independently gauge invariant ; \hfill\break
  \indent 3) $m(1,...,n) = (-1)^n m(n,...,1)$ ; \hfill\break
  \indent 4) they satisfy the ``dual Ward identity,'' which for $n=4$ is
\beq
m(1,2,3,4)+m(2,1,3,4)+m(2,3,1,4)=0  ;
\label{eq:ward}
\eeq
  \indent 5) they factorize in the soft gluon limit and in
    the limit in which two of \hfill\break
\indent the gluons are collinear; \hfill\break
 \indent 6) they are incoherent (to leading order in the number of colors,
 in general, and completely for $n=4$); for $n=4$ one finds
\beq 
\sum_{\rm colors} |\M|^2 = {g^2 A^2 \o 4} N^2 (N^2-1) 
\sum_{\rm perms} |m(1,2,3,4)|^2. 
\label{eq:incoh}
\eeq
 \indent (See Appendix A for the proof.) 

The complete set of sub-amplitudes can be obtained from the following three
\cite{hgggg}:
\beqa
&&m(1^+,2^+,3^+,4^+)={\mh^4\over \la 1 2\ra \la 2 3\ra
   \la 3 4\ra \la 4 1\ra} 
\label{eq:pppp}\\ 
&&m(1^-,2^+,3^+,4^+) = -
{\la 1{-}|\slash \ph |3{-} \ra^2 [24]^2 \over S_{124} S_{12} S_{14}}     
-{\la 1{-}|\slash \ph | 4{-} \ra^2 [2 3]^2 \over S_{123} S_{12} S_{23}} \no\\
&&\qq
-{\la 1{-}|\slash \ph | 2{-} \ra^2 [3 4]^2 \over S_{134} S_{14} S_{34}} 
 +{[2 4] \over 
[ 1 2 ] \la 2 3\ra \la 3 4 \ra   [ 4 1]} 
\biggl\{ S_{23} {\la 1{-} |\slash \ph | 2{-} \ra \over \la 4 1\ra } \no \\
&&\qq\qq\qq\qq\qq\qq\qq +       S_{34} 
      {\la 1{-} |\slash \ph | 4{-} \ra \over \la 1 2\ra }
-[2 4 ] S_{234}\biggr\} 
\label{eq:mppp}\\ 
&&m(1^-,2^-,3^+,4^+)=-{\la 1 2\ra^4\over \la 12 \ra 
\la 23 \ra \la 34\ra \la 41\ra}
 -{[34]^4 \over [1 2] [23] [34] [41]} .
\label{eq:mmpp}
\eeqa
The structures containing $\ph$ can be expanded in terms of spinor products 
using Eq.~\ref{eq:helproj} and momentum conservation.  For example 
$\la 1{-}|\slash \ph |3{-} \ra = - ( \la 12 \ra [23] + \la 14 \ra [43] )$.
Permutations of $m(1^+,2^+,3^+,4^+)$ are obtained by permuting the 
momenta in the right side of Eq.~\ref{eq:pppp} identically. Permutations of
$m(1^-,2^+,3^+,4^+)$ are obtained by permuting $p_2$, $p_3$ and $p_4$
in the right side of Eq.~\ref{eq:mppp} then using 
the cyclic and reversal properties
of the sub-amplitudes.  Permutations of
$m(1^-,2^-,3^+,4^+)$ are obtained by permuting the momenta in the 
{\it denominators} of the right side of Eq.~\ref{eq:mmpp} only. 
It is straightforward to
check that the sub-amplitudes obtained from Eqs.~\ref{eq:pppp}-\ref{eq:mmpp}
in this way
obey the requisite relations.   
The amplitudes for the other helicity combinations can be
obtained (modulo phases) by parity transformations.  

\section{The Higgs Boson Plus Quark Anti-quark and Two Gluon Amplitude}
The $H  q \bar q g g$ amplitude 
can be obtained from the Feynman diagrams
of Fig. 5.  As was the case for the $Hgggg$ amplitude, 
the calculation can be simplified by  judicious choice of color decomposition
\cite{kunszt,parkii}.
The amplitude for a Higgs boson, a quark--anti-quark pair with color indices
$i,j$ and $n$ gluons with color indices $a_1,...,a_n$ can be 
written:
\beq
\M = -i g^n A
\sum_{\rm perms} (T^{a_1} T^{a_2} ... T^{a_n})_{ij} m(p1,\e_1;...;p_n,\e_n),
\label{eq:qdual}
\eeq
where the sum runs over all $n!$ permutations of the gluons and the 
sub-amplitudes $m(p1,\e_1;...;p_n,\e_n)$ have an implicit dependence on
the momenta and helicities of the quark and anti-quark.  For the case we
are interested in there are only two subamplitudes which we will label as
$m(3,4)$ and $m(4,3)$ since the gluon momenta are $p_3$ and $p_4$.
Like the subamplitudes for the pure gluon case these subamplitudes are
separately gauge independent and factorize in the soft gluon and collinear
particle limits.

As in the case of the $Hq\bar q g$ amplitude the quark and anti-quark must have
opposite helicities.
Labelling the helicity amplitudes by the helicity of the quark, anti-quark
and the two gluons (in that order) we find
\beqa
m^{+-++}(3,4) &=& { \la 2{-}|\slash \ph| 3{-} \ra^2 \o S_{124} }
                 { [14] \o \la 24 \ra } \l({1\o S_{12}} + {1\o S_{14}} \r)
            -{ \la 2{-}|\slash \ph| 4{-} \ra^2 \o S_{123} S_{12}}
             { [13] \o \la 23 \ra }
\nonumber \\
            &+&{ \la 2{-}|\slash \ph| 1{-} \ra^2 \o [12] 
            \la 23 \ra \la 24 \ra \la 34 \ra }
\label{eq:qpmpp}
\eeqa
To get the subamplitude with the other ordering, $m^{+-++}(4,3)$, exchange
$p_3 \leftrightarrow p_4$ in this expression.  The other independent 
subamplitudes are
\beqa
m^{+-+-}(3,4) &=& -{\la24\ra^3 \o \la12\ra \la23\ra \la34\ra } 
                 + { [13]^3 \o [12][14][34] }
\label{eq:qpmpmi} \\
m^{+-+-}(4,3) &=& -{ [13]^2 [23] \o [12][24][34] }
                  + {\la14\ra \la24\ra^2 \o \la12\ra \la13\ra \la34\ra } 
\label{eq:qpmpmii} 
\eeqa
The other helicity amplitudes (up to phases) can be obtained by parity $(P)$, 
Bose symmetry $(B)$
and charge conjugation $(C)$ transformations: 
\beqa
|\M^{-+--}|^2 \pe |\M^{+-++}|^2,~~ 
|\M^{-+++}|^2 \ce |\M^{+-++}|^2_{1\ch2},
\nonumber \\
|\M^{+---}|^2 \cpe |\M^{+-++}|^2_{1\ch2},~~ 
|\M^{-+-+}|^2 \pe |\M^{+-+-}|^2,
\nonumber \\
|\M^{+--+}|^2 \be |\M^{+-+-}|^2_{3\ch4},~~
|\M^{-++-}|^2 \bpe |\M^{+-+-}|^2_{3\ch4}.
\label{eq:qsym} 
\eeqa

\section{The Higgs Boson Plus Two Quark AntiQuark Pair Amplitude}

The remaining processes producing a Higgs boson plus two jets are those
involving a combination of four quarks and anti-quarks.  In the case where
the two pairs are of different flavors the amplitude can be obtained
from the Feynman diagram in Fig. 6.  In the case when the two pairs are
identical there is an additional diagram which can be obtained by switching
the $2 \ch 4$ in the diagram of Fig. 6.  We present the amplitude for the case
of two different quark pairs, since the identical case can be obtained from it.
The sole independent
helicity amplitude can be labelled in terms of the helicities of the 1st
quark, the 1st antiquark, the 2nd quark and the 2nd antiquark (in that order):
\beq
\M^{+-+-} = iA g^2 T^a_{ij}T^a_{kl} 
\l( {\la24\ra^2 \o \la12\ra \la34\ra} + { [13]^2 \o [12][34] } \r).
\label{eq:hqqqq}
\eeq
The other helicity amplitudes can be obtained by parity and charge conjugation
transformations:
\beqa
|\M^{-+-+}|^2 &\pe& |\M^{+-+-}|^2,~~
|\M^{-++-}|^2 \ce |\M^{+-+-}|_{1\ch2}^2, \nonumber \\ 
|\M^{+--+}|^2 &\ce& |\M^{+-+-}|_{3\ch4}^2.
\label{eq:qqqqsym}
\eeqa
\section{The Soft Higgs Limit}

The effective Lagrangian, Eq. \ref{eq:leff}, implies that for the case of 
constant Higgs field $H$, {\it i.e.}, a Higgs boson with no momentum,
the amplitude for a process containing a Higgs boson reduces to the amplitude
for the process without the Higgs boson times an overall factor of $A$.
For the $Hgggg$ amplitude only the helicity conserving ${+}{+}{-}{-}$ 
amplitude 
survives in the $\ph \to 0$ limit.  In this limit one can show that the 
two terms in Eq. \ref{eq:mmpp} are equal, regardless of the ordering of the
momenta.  As an example, consider the subamplitude
\beq
m(1^-,3^+,2^-,4+)=-{\la 1 2\ra^4\over \la 13 \ra 
\la 32 \ra \la 24\ra \la 41\ra}
 -{[34]^4 \over [1 3] [32] [24] [41]}.
\label{eq:mmppl}
\eeq
In the limit $\ph\to 0$ the second term in this expression becomes
\beqa
{[34]^4 \la31\ra \la23\ra \la42\ra \la14\ra
               \o S_{13}S_{32}S_{24}S_{41} } &=&
{\la 3{+}| \slash p_4 | 2{+} \ra  \la 3{+}| \slash p_4 | 1{+} \ra
 \la 4{+}| \slash p_3 | 1{+} \ra  \la 4{+}| \slash p_3 | 2{+} \ra 
               \o S_{13}S_{32}S_{24}S_{41} } \nonumber \\
&=&{\la 3{+}| \slash p_1 | 2{+} \ra  \la 3{+}| \slash p_2 | 1{+} \ra
 \la 4{+}| \slash p_2 | 1{+} \ra  \la 4{+}| \slash p_1 | 2{+} \ra 
               \o S_{13}S_{32}S_{24}S_{41} } \nonumber \\
&=&{ \la12\ra^4 [13][32][24][41] \o  S_{13}S_{32}S_{24}S_{41} },
\label{eq:secondt}
\eeqa
where momentum conservation as used
in the second line. 
When common factors are cancelled Eq.~\ref{eq:secondt} is identical to the
first term of Eq. \ref{eq:mmppl}.  The final result
agrees with the well-known
form of the pure gluon subamplitudes\cite{parki,parkii,parkiii}.

The $Hq\bar q gg$ subamplitudes reduce to the $q\bar q gg$ subamplitudes:
\beqa
m^{+-+-}(3,4) &\plim& - {2 [13]^3 \o [12][14][34] }
\label{eq:qqlimi} \\
m^{+-+-}(4,3) &\plim&  - {2\la14\ra \la24\ra^2 \o \la12\ra \la13\ra \la34\ra } 
\label{eq:qqlimii} 
\eeqa
Finally, the soft-Higgs limit of the $Hq \bar q q^\prime \bar q^\prime$
amplitude is
\beq
\M^{+-+-} \plim  -2iAg^2 T^a_{ij}T^a_{kl} 
{\la24\ra^2 \o \la12\ra \la34\ra},
\label{eq:hqqqqlim}
\eeq
which has the proper relation to the amplitude for 
$q\bar q q^\prime \bar q^\prime$.

\section{Factorization of the Amplitudes}
The helicity amplitudes we have calculated factorize in the limit that
a gluon becomes soft or two particles become collinear.  This property has
been established for the pure QCD processes involving quarks and gluons.
We will present some representative examples of the factorization limits
of our amplitudes.  

The simplest cases involve the reduction of the Higgs plus 3 particle 
amplitudes to the Higgs plus two gluon amplitude in the appropriate
limit.  Discussion of these limits is facilitated by expressing the 
$Hgg$, $Hggg$, and $Hq\bar qg$ amplitudes in the same dual color
decompositions as we used for the $Hgggg$ and $Hq\bar qgg$ amplitudes.
For the $Hgg$ case there is only one subamplitude, $m_{gg}(1,2)$, which 
can be obtained from Eq.~\ref{eq:hggamp} by replacing $\delta_{ab}$ by 1.
For the $Hggg$ case there are two subamplitudes but the Ward identity
ensures that they are equal and opposite.  The subamplitude $m_{ggg}(1,2,3)$
can be obtained from Eq.~\ref{eq:hgggamp} by replacing $f_{abc}$ by $-i$.
The lone subamplitude for $Hq\bar qg$ is Eq.~\ref{eq:hqqgamp} with the 
factor of $T^a_{ij}$ removed.  

Taking one of the gluons to be soft in $m_{ggg}(1^+,2^+,3^+)$ yields
\beq
m_{ggg}(1^+,2^+,3^+) \stackrel{p_1\to0}{\to} 
-\l\{ {g\sqrt{2}\la32\ra \o \la12\ra \la31\ra } \r\}m_{gg}(2^+,3^+),
\label{eq:mgggsoft}
\eeq
where the factor in brackets is the square root of the ``eikonal factor''.
This is in keeping (up to phase conventions)
with the general result of Mangano {\it et al.} 
\cite{parki}.
A similar limit applies to $m_{ggg}(1^-,2^+,3^+)$.  Taking two of the gluons to
be collinear is accomplished by letting $p_1 \to zP$ and $p_2 \to (1-z)P$:
\beqa
m_{ggg}(1^+,2^+,3^+) &\to& 
\l( {ig\sqrt{2}[12] \o \sqrt{z(1-z)} } \r) \l({-i\o S_{12}} \r) 
m_{gg}(P^+,3^+), \nonumber \\
m_{ggg}(1^-,2^+,3^+) &\to& 
\l( {ig\sqrt{2}\la12\ra(1-z)^2 \o \sqrt{z(1-z)} } \r) \l({-i\o S_{12}} \r) 
m_{gg}(P^+,3^+),
\label{eq:mgggcoll}
\eeqa
which is again consistent with the general result\cite{parki}.

The $Hgggg$ subamplitudes exhibit the same factorization properties.
Notice that only the momenta which are adjacent in the argument of the 
subamplitude appear paired in the denominators.
For example, taking $p_1$ to be soft in $m(1^+,2^+,3^+,4^+)$ gives
\beq
m(1^+,2^+,3^+,4^+) \stackrel{p_1\to0}{\to} 
{ S_{234}^2 \o \la12\ra \la23\ra \la34\ra \la41\ra }
= { \sqrt{2} \la42\ra \o \la12\ra \la41\ra } m_{ggg}(2^+,3^+,4^+),
\label{eq:mggggsoft}
\eeq
in agreement with the general result.  The ${\cal O}(1/S_{ij})$ singularities
in Eq.~\ref{eq:mppp}
are in reality only ${\cal O}(1/\sqrt{S_{ij}})$ singularites.  For example,
taking $p_1 \to zP$ and $p_2 \to (1-z)P$ in $m(1^-,2^+,3^+,4^+)$ gives
\beqa
&&m(1^-,2^+,3^+,4^+) \stackrel{1\parallel2}{\to}
{1\o\sqrt{z(1-z)}} \l \{ {(1-z)^2 \o [12] } 
{ S_{P34}^2 \o \la P3\ra \la34\ra \la4P\ra }
- { z^2 \o \la12\ra} {[34]^3 \o [P3][P4] } \r\} \nonumber \\
&& = {\sqrt{2}\o\sqrt{z(1-z)}} \l\{ {(1-z)^2 \o [12] } m_{ggg}(P^+,3^+,4^+)
- { z^2 \o \la12\ra}  m_{ggg}(P^-,3^+,4^+) \r \}. \no \\
\label{eq:mpppcoll}
\eeqa

The $Hq\bar qgg$ and $H q\bar q q^\prime \bar q^\prime$ amplitudes also factor.
Taking the limit of $p_1 \parallel p_2$ as before gives:
\beqa
m^{+-++}_{q\bar q gg} &\stackrel{1\parallel2}{\to}&
{(1-z)\o [12]} { S_{P34} \o \la P3\ra \la34\ra \la4P\ra }
+{z\o \la12\ra } { [34]^3 \o [P3] [P4] } 
\nonumber \\
&=&
{(1-z)\o [12]}  m_{ggg}(P^+,3^+,4^+)
+ {z\o \la12\ra } m_{ggg}(P^-,3^+,4^+) 
\label{eq:mqqggcoll}
\eeqa
and
\beq
M^{+-+-}_{q\bar q q^\prime\bar q^\prime} \stackrel{1\parallel2}{\to}
-i g^2 T^a_{ab}T^a_{dc} \l\{ {(1-z)\o [12]} m_{q\bar qg}^{+-+}
+ {z \o \la12\ra } m_{q\bar qg}^{+--} \r \}.
\label{eq:mqqqqcoll}
\eeq

The $Hgggg$ subamplitudes also factor in the three-gluon channel.  Letting
$P = p_1 + p_2 + p_3$ and taking $P^2 \to 0$ in the ${-}{+}{+}{+}$ subamplitude
gives 
\beq 
m(1^-,2^+,3^+,4^+) \to [P4]^2 {1\o P^2} {\la 1P\ra^2 [23]^2 \o S_{12} S_{23} }
\sim m_{gg}(4+,P+) {1\o P^2} \tilde m(1^-,2^+,3^+,P^-),
\label{eq:threepole}
\eeq
where $\tilde m$ is the four gluon subamplitude (without a Higgs boson)
and the $\sim$ indicates equality modulo phases.  Identical relations 
hold for the $1/S_{124}$ and $1/S_{134}$ poles.  Since the four-gluon 
amplitude is helicity conserving the other helicity amplitudes have no
three-gluon poles.  Likewise there is no $1/S_{234}$ pole in 
$m(1^-,2^+,3^+,4^+)$.

\section{Numerical Results and Conclusions}

We will present numerical results for the CERN Large Hadron Collider (LHC)
at a center-of-mass energy of $\sqrt{S}=14~\TeV$ and the Fermilab Tevatron
at $\sqrt{S}=2~\TeV$.
Since all the parton level cross sections 
are singular in the small $\pt$ limit of one of the 
jets we will place a $\pt$ cut on the outgoing jets.  Since there are also
collinear singularities we will require that the outgoing jets be separated 
by $\Delta R_{ij}\equiv \sqrt{\Delta \phi_{ij}^2+\Delta \eta_{ij}} \ge 0.7$. 
We will also require the outgoing jets have rapidity $\mid y\mid <2.5$.
Since there are no singularities depending on the momentum of the Higgs boson
we will allow it to be unconstrained, except for a $\pt$ cut.

The results separated according to parton processes are presented in 
Fig.'s 7 and 8.  We see that at the LHC the all gluon process dominates
as expected with the $qg \to qgH$ process and its charge conjugate contributing
an additional 15\%.  The other processes are negligible.  At the Tevatron
the pure gluon process and the $qg \to qgH$ process give roughly equal
contributions.  

Fig.'s 9 and 10 show the result of varying the transverse momentum cut (on the
jets and the Higgs boson simultaneously).  We see that in both cases the
cross section drops sharply with the increasing $\pt$ cut.  The dependence
of the cross section on the other cuts is weak.  Increasing the minimum
$\Delta R$ to 1.0 decreases the cross section by about 15\%.  Requiring that
the jets be separated from the Higgs boson the same $\Delta R$ also reduces
the cross section be about 15\%.

In summary, we have presented the amplitudes for the production of a Higgs
accompanied by two jets.  We find that the cross section is around a few 
picobarns at the LHC and a few hundredths of a pb at the Tevatron.  
Our results provide the ``real'' corrections to Higgs production at non-zero
transverse momentum.  They can be combined with the virtual corrections
to complete the next-to-leading order calculation.

\pagebreak

\section{Appendix. Squaring the Amplitudes}

We first proceed to verify the incoherence of the subamplitudes for $ggggH$.
Using the fact that subamplitudes are invariant (for $n=4$) under reversal
of the order of the arguments we write
\beqa
\M &=& \Big\{ \l[ \tr(T^a T^b T^c T^d) + \tr(T^d T^c T^b T^a) \r] m(1,2,3,4)
\nonumber \\   
&+& \l[ \tr(T^a T^b T^d T^d) + \tr(T^d T^d T^b T^a) \r] m(1,2,4,3)
\nonumber \\   
   &+& \l[ \tr(T^a T^c T^b T^d) + \tr(T^d T^b T^c T^a) \r] m(1,3,2,4) \Big\}.
\label{eq:mexpl}
\eeqa
It is straightforward to show that the squared color factors are
\beqa
c_1 &=& \sum_{\rm colors} \l[ \tr(T^a T^b T^c T^d) + \tr(T^d T^c T^b T^a) \r]^2
\nonumber \\
&=& { (N^2-1)^2 \o 4N^2} + {1\o16}\l[ \l({5\o3}\r)^2 (N^2-1) + N^2(N^2-1) \r]
\label{eq:csquare}
\eeqa
and that the cross terms are
\beqa
c_2 &=&
 \sum_{\rm colors} \l[ \tr(T^a T^b T^c T^d) + \tr(T^d T^c T^b T^a)^* \r]
                   \l[ \tr(T^a T^b T^d T^c) + \tr(T^c T^d T^b T^a)^* \r]
\nonumber \\
&=& { (N^2-1)^2 \o 4N^2} + {1\o16}\l[ \l({5\o3}\r)^2 (N^2-1) - N^2(N^2-1) \r]
\label{eq:ccross}
\eeqa
The amplitude squared is then
\beqa
\sum_{\rm colors} |\M|^2&=& 
4g^4A^2 \Big\{ c_1 \big[ |m(1,2,3,4)|^2 + |m(1,2,4,3)|^2 +
                                       |m(1,3,2,4)|^2 \big] \nonumber \\
\quad\quad\quad
      &+& c_2 \big[ m(1,2,3,4)( m(1,2,4,3)^* + m(1,3,2,4)^*) \nonumber \\
\quad\quad\quad &+& m(1,2,4,3)( m(1,2,3,4)^* + m(1,3,2,4)^*) \nonumber \\
 \quad\quad\quad  &+&  m(1,3,2,4)( m(1,2,3,4)^* + m(1,2,4,3)^*) \big] \Big \}.
\label{eq:tampsq}
\eeqa
Use of the Ward identity, Eq. (\ref{eq:ward}),
 allows all the cross terms to be written as subamplitudes squared.  When this
is done and the above results for the color factor are used, 
Eq. (\ref{eq:incoh}) follows.

The following identities were used to relate the spinor products to traces 
over gamma matrices\cite{parki}:
%
\beqa
[ i_1 i_2 ] \la i_2 i_3 \ra...\la i_{2n} i_1 \ra &=&
 \{ i_1 i_2... i_{2n} P_+ \}
\no \\
\la i_1 i_2 \ra [ i_2 i_3 ]...[ i_{2n} i_1] &=& \{ i_1 i_2... i_{2n} P_- \},
\label{eq:trace}
\eeqa
where $P_\pm={1\o2}(1\pm\gamma_5)$, and $\{i_1 i_2... i_n\}$ denotes the trace
of $\slash p_1 \slash p_2... \slash p_n$.  In order to reduce the traces which 
contained a factor of $\g_5$, we use the identity\cite{parki}:
%
\beqa
&&\{i_1 i_2... i_{2n}\gamma_5\} \{j_1 j_2... j_{2m}\gamma_5\} =
 \{i_1 i_2... i_{2n}\} \{j_1 j_2... j_{2m}\} \no \\
&&\qq\qq\qq - 2\Big[ [ i_1 i_2 ] \la i_2 i_3 \ra...\la i_{2n} i_1 \ra
\la j_1 j_2 \ra [ j_2 j_3]...[ j_{2n} j_1 ] + c.c.\Big].
\label{eq:trproduct}
\eeqa

We first consider the square of the $ggggH$ subamplitudes.
For the case of $m(1^+,2^+,3^+,4^+)$, we have
\beq
| m(1^+,2^+,3^+,4^+) |^2 = {\mh^8 \o S_{12} S_{23} S_{34} S_{41}}.
\label{eq:g4ppppsq} 
\eeq
In the case of $m(1^-,2^+,3^+,4^+)$, we rewrite Eq.~\ref{eq:mppp} in a
more compact form:
\beqa
&&m(1^-,2^+,3^+,4^+) = -
{\la 1{-}|\slash \ph |3{-} \ra^2 [24]^2 \o S_{124} S_{12} S_{14}}     
-{\la 1{-}|\slash \ph | 4{-} \ra^2 [2 3]^2 \o S_{123} S_{12} S_{23}} \no\\
&&
-{\la 1{-}|\slash \ph | 2{-} \ra^2 [3 4]^2 \o S_{134} S_{14} S_{34}} 
 +{[2 4] \o  [ 1 2 ] [14]  \la 1 3\ra } 
\biggl\{  {\la 1{-} |\slash \ph | 2{-} \ra^2 \o \la14\ra \la34\ra }
+  {\la 1{-} |\slash \ph | 4{-} \ra^2 \over \la12\ra \la23\ra} \biggr\}. \qq
\label{eq:mpppalt}
\eeqa
We then write
\beq
| m(1^-,2^+,3^+,4^+) |^2 = \sum_{\rm i=1}^{\rm 5} \sum_{\rm j=1}^{\rm i}
 {\rm {n_{ij}\o d_i d_j} },
\label{eq:g4mpppsq}
\eeq
where the independent terms are:
\beqa
n_{11} &=& {1 \o 4} S_{24}^2\{1(2{+}4)3(2{+}4)\}^2, \qq
n_{44} = {1\o4} S_{12}S_{13}S_{24}S_{34} \{1(3{+}4)2(3{+}4)\}^2 \no \\
n_{12} &=& \{1(2{+}4)324(2{+}3)\}^2 -
2  S_{23} S_{24}(S_{12} S_{24} {+} S_{13} S_{34} + \{1243\}) \no \\
&&\qq \qq \qq \qq \qq \qq \,\, \, \times
(S_{12}S_{23} {+} S_{14}S_{34} {+} \{1234\})  \no \\
n_{14} &=& -S_{24} \Big[ \{1(2{+}4)342(3{+}4)\} \{1(2{+}4)312(3{+}4)\} \no \\
&&\qq -\{1243\}(S_{12}S_{23}{+}S_{14}S_{34}{+}\{1234\})
(S_{13}S_{23}{+}S_{14}S_{24}{+}\{1324\}) \Big] \no \\
-4n_{24} &=&
\{1(2{+}3)432(3{+}4)\}\{1(2{+}3)42\}\{132(3{+}4)\} \no \\
&& -\{1(2{+}3)432(3{+}4)\}(\{1234\}^2{-}4S_{12}S_{23}S_{34}S_{41}) \no \\
&& -\Big[\{1(2{+}3)432(3{+}4)\}\{1234\}\no \\
&&\;\; +2S_{23}S_{34}(S_{12}\{1324\} {-}S_{13}\{1234\}
{+}S_{14}\{1243\} {+} 2S_{12}S_{14}S_{24}) \Big] \no \\
&& \;\;\times
(\{1(2{+}3)42\} {-} \{132(3{+}4)\}) \no\\
n_{25} &=& - {1 \o 4} S_{23} \{1(2{+}3)4(2{+}3)\}^2\{1324\} \no \\
n_{45} &=& -S_{13}S_{24}\Big[\{1(3{+}4)234(2{+}3)\}\{1(3{+}4)214(2{+}3)\}
\no \\
&&\qq - \{1234\} (S_{13}S_{23} {+} S_{14}S_{24} {+} \{1324\})
       (S_{12}S_{24} {+} S_{13}S_{34} {+} \{1243\})
\Big]
\no \\
d_1 &=& S_{12} S_{14} S_{124}, \qq d_4 = S_{12} S_{13} S_{14} S_{34}.
\label{eq:paradigm}
\eeqa
The remaining terms can be obtained by switching the momenta:
\beqa
n_{22} &=& n_{11}(3{\ch} 4), \qq
n_{33} = n_{11}(2{\ch} 3), \qq
n_{55} = n_{44}(2{\ch} 4), \no \\
n_{13} &=& n_{12}(2{\ch} 4), \qq
n_{15} = n_{14}(2{\ch} 4), \qq
n_{23} = n_{13}(3{\ch} 4), \no \\
n_{34} &=& n_{25}(2{\ch} 4), \qq
n_{35} = n_{24}(2{\ch} 4), \qq
d_5 = d_4(2{\ch} 4),\no \\
d_2 &=& d_1(1{\ch} 2, 3{\ch} 4), \qq
d_3 = d_1(1{\ch} 4, 2{\ch} 3). \no \\
\label{eq:switches}
\eeqa
The two independent permutations of $m(1^-,2^-,3^+,4^+)$ squared are:
\beqa
|m(1^-,2^-,3^+,4^+)|^2 &=& {S_{12}^3 \o S_{14} S_{23} S_{34}} 
			  +{S_{34}^3 \o S_{12} S_{14} S_{23}}
			  +{\{1234\}^2 - 2S_{12}S_{23}S_{34}S_{41}
			    \o S_{14}^2 S_{23}^2} \no \\
|m(1^-,3^+,2^-,4^+)|^2 &=& {S_{12}^4 + S_{34}^4 \o S_{13}S_{14}S_{23}S_{34}}
\no \\
&&+ \Big[ (\{1234\}^2 {-} 2S_{12}S_{23}S_{34}S_{41})
     (\{1243\}^2 {-} 2S_{12}S_{24}S_{43}S_{31}) \no \\
&&\;\; +\{1234\}\{1243\}(\{1234\}\{1243\} {+}2S_{12}S_{34}\{1324\}) \Big] \no 
\\
&&  / [2(S_{13}S_{14}S_{23}S_{34})^2]. 
\label{eq:g4mmppsq}
\eeqa

Turning our attention to the $q \bar q gg H$ amplitudes,  we break the 
subamplitude $m^{{+}{-}{+}{+}}(3,4)$ into a symmetric piece, 
$m_s$, and anti-symmetric piece, $m_a$,
under interchange of the two gluons.  
The squared amplitude is then given by
\beq
|\M^{{+}{-}{+}{+}}|^2 = 2 A^2 g^4
\l[ C_1 (|m_{s}|^2 + |m_{a}|^2) + C_2 (|m_{s}|^2 
- |m_{a}|^2) \r],
\label{eq:qqggpmpp}
\eeq 
where the color factors are 
$C_1={(N^2-1)^2\over4N}$ and $C_2=-{(N^2-1)\over4N}$.
We write:
\beqa
|m_s|^2 &=&{1\over4} \l({n_{11}\over d_1^2} + {n_{22}\over d_2^2} 
+ {n_{12}\over d_1 d_2}\r)\\
|m_a|^2 &=&{n_{11}\over d_3^2} + {n_{22}\over d_4^2} + {n_{33}\over d_5^2} 
- {n_{12}\over d_3 d_4} + {n_{13}\over d_3 d_5} + {n_{23}\over d_4 d_5}
\label{eq:symmasymm}
\eeqa
where the numerator and denominator terms are	
\beqa 
&&n_{11} = {1\over4}{S_{14} S_{24}}\{2(1{+}4)3(1{+}4)\}^2,\qquad
n_{22} = n_{11}{(3\ch4)},\no\\
&&n_{33} = {1\over4}{S_{12} S_{23} S_{24} S_{34}}\{1(3{+}4)2(3{+}4)\}^2,\no\\
&&n_{12} = \{2(1{+}4)314(1{+}3)\}\{2(1{+}4)324(1{+}3)\}\no\\
&&\qquad- \{1324\}(S_{12} S_{13} + S_{24} S_{34} + \{1243\}) 
(S_{12} S_{14} + S_{23} S_{34} + \{1234\})\no\\
&&n_{13} = -S_{24} n_{12}{(2\ch4)},\qquad
n_{23} = -S_{23} n_{13}{(3\ch4)},\no\\
&&d_1 = S_{14} S_{24} S_{124},\qquad
d_2 = S_{13} S_{23} S_{123},\qquad
d_5 =  S_{12} S_{23} S_{24} S_{34},\no\\
&&d_4 = S_{23} S_{123}\l({1\over S_{12}}
+{1\over2S_{13}}\r)^{-1},\qquad
d_3 = S_{24} S_{124}\l({1\over S_{12}}+{1\over2S_{14}}\r)^{-1}.\no\\
\label{eq:qqgg1sub}
\eeqa
For the other independent helicity amplitude, squaring yields 
\beqa
&&|\M^{{+}{-}{+}{-}}|^2 = g^4 A^2 \{C_1 
(|m^{{+}{-}{+}{-}}(3,4)|^2 + |m^{{+}{-}{+}{-}}(4,3)|^2) \no \\
&& \qq\qq\qq\qq +
2C_2 \re[m^{{+}{-}{+}{-}}(3,4)m^{{+}{-}{+}{-}}(4,3)^*]\},
\label{eq:qqggpmpm}
\eeqa
where
\beqa
\lefteqn{
|m^{{+}{-}{+}{-}}(3,4)|^2 = \biggl\{{S_{13}^3\over S_{12} S_{14} S_{34}} 
+ {S_{24}^3\over S_{12} S_{23} S_{34}}
+ {1\over S_{14} S_{23} S_{12}^2 S_{34}^2} }  \no\\
&&\qq\times\Big[-\{1243\}^2 \{1324\} - S_{13} S_{24} \{1234\} \{1243\} 
+ S_{12} S_{13} S_{24} S_{34} \{1324\}\Big] \biggr\},\no\\
\lefteqn{|m^{{+}{-}{+}{-}}(4,3)|^2 = 
\biggl\{{S_{13}^2 S_{23}\over S_{12} S_{24} S_{34}} 
+ {S_{14} S_{24}^2 \over S_{12} S_{13} S_{34}} + { \l(\{1243\} \{1234\} 
+ S_{12} S_{34} \{1324\}\r) \over S_{12}^2 S_{34}^2 } \biggr\},} \no\\
\lefteqn{2\re[m^{{+}{-}{+}{-}}(3,4)m^{{+}{-}{+}{-}}(4,3)^*]
= -{\{1324\}\over S_{12} S_{34}} 
\l({S_{13}^2\over S_{14} S_{24}}
+ {S_{24}^2\over S_{13} S_{23}}\r). }
\label{eq:qqgg2sub}
\eeqa

For the $q \bar q q^\prime\bar q^\prime$ amplitude, the square of 
Eq.~\ref{eq:hqqqq} yields
\beq
\M^2 = {A^2 g^4 (N^2-1)\over 4 S_{12} S_{34}} \l[(S_{13} - S_{24})^2 
+ {\{1243\}^2\over S_{12} S_{34}}\r]. 
\label{eq:hqqqqsq}
\eeq
In the case of identical quark pairs, there is a second diagram whose square
can be obtained by switching $1\ch 3$ in Eq.~\ref{eq:hqqqqsq}.  The 
interference term which arises is
\beq
-2\re[\M \M^*(1{\ch}3)] = {-A^2g^4 (N^2{-}1) \o 4N}
\l[{(S_{13} - S_{24})^2 \{1234\} 
- 2\{1324\} \{1243\}\over S_{12} S_{23} S_{14} S_{34}}\r].
\label{eq:hqqqqint}
\eeq

\bigskip
\begin{center}
{\bf ACKNOWLEDGMENTS}
\end{center}
\medskip
The author's would like to thank S. Dawson for her work on the initial
stages of this project and for helpful discussions.
S.D. and D.R. would like to thank the Hackman Scholar program at Franklin
and Marshall College for financial support.

\pagebreak

\centerline{\bf FIGURE CAPTIONS}
\bigskip

\noi {\bf Figure 1}.  The vertices and Feynman rules of the effective theory.  The
curly lines indicate gluons and the dashed lines indicate the Higgs boson.
\medskip

\noi {\bf Figure 2}.  The Feynman diagrams for the $gggH$ amplitude.  There are
two more diagrams of the same form as b), where the Higgs boson attaches to 
gluon 1 and gluon 3.
\medskip

\noi {\bf Figure 3}.  The Feynman diagram for the $q\bar qgH$ amplitude.

\noi {\bf Figure 4}.  The Feynman diagrams for the $ggggH$ amplitude.  
There are
12 diagrams of type a), 3 of type b), 1 of type c), 4 of type d), and 6 of
type e), for a total of 26.
\medskip

\noi {\bf Figure 5}.  The Feynman diagrams for the $q\bar q ggH$ amplitude.  There
is one diagram of type a), two of type b), 4 of type c) and one of type d),
for a total of 8.
\medskip

\noi {\bf Figure 6}.  The Feynman diagram for the $q\bar q q^\prime \bar q^\prime H$
amplitude.  In the case when the quark pairs are identical there is a second
diagram with the quark lines switched.
\medskip

\noi {\bf Figure 7}.  The various contributions to the cross section for production 
of a Higgs boson plus two jets at the LHC as a function of the mass of the 
Higgs boson.  A $\pt$ cut of 50 GeV has been 
placed on the jets and the Higgs boson.  The labels are as follows:
 $gggg$ represents $gg\to ggH$;  $qgqg$ represents $qg\to qgH$ and 
its complex conjugate; $ggqq$ represents 
$gg\to q\bar q H$; $qqqq$ represents all the processes involving
two incoming quarks or antiquarks and two outgoing quarks or antiquarks;
$ggqq$ represents $gg\to q\bar qH$.
\medskip

\noi {\bf Figure 8}.  The various contributions to the cross section for production 
of a Higgs boson plus two jets at the Tevatron as a function of the mass of 
the Higgs boson. 
A $\pt$ cut of 25 GeV has been 
placed on the jets and the Higgs boson.  The labels are the same as in 
Fig. 7.
\medskip

\noi {\bf Figure 9}.  The cross section for production of a Higgs boson plus
two jets at the LHC for three values of the $\pt$ cut.
\medskip

\noi {\bf Figure 10}.  The cross section for production of a Higgs boson plus
two jets at the Tevatron for three values of the $\pt$ cut.

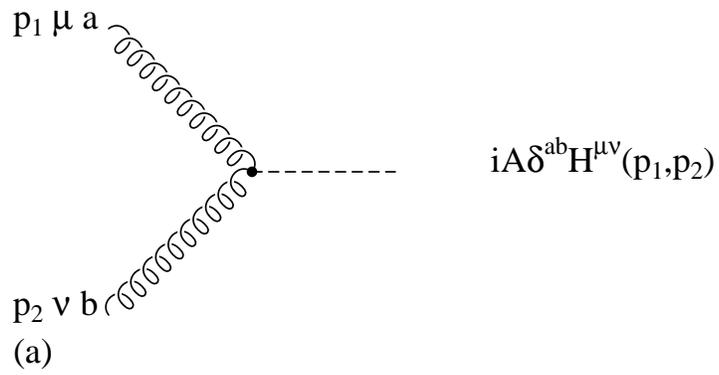
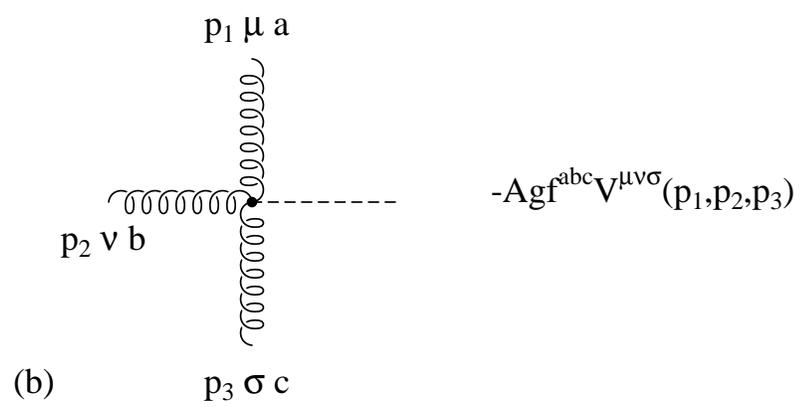
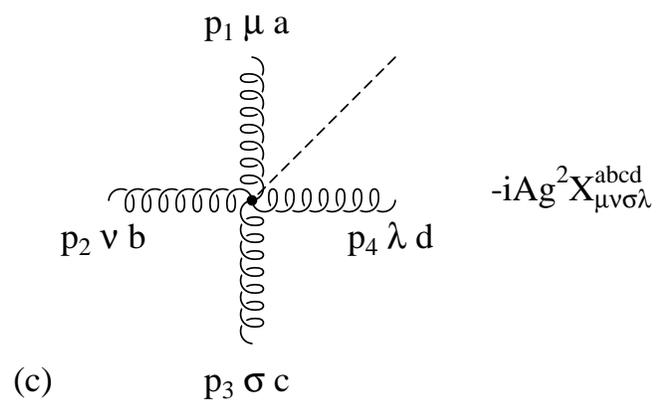
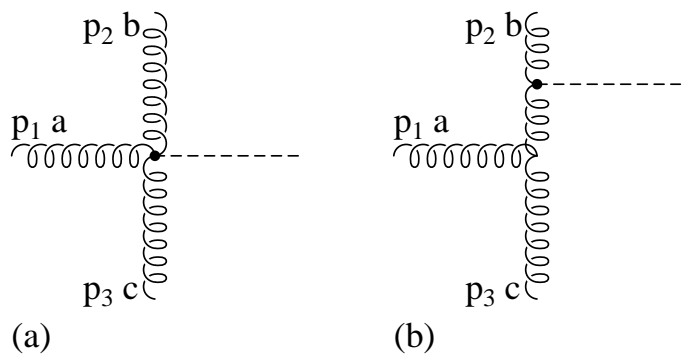
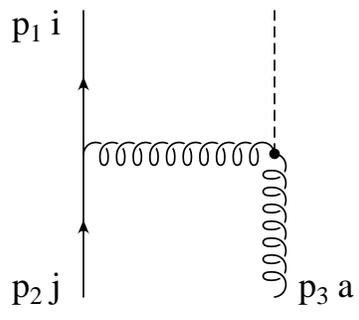
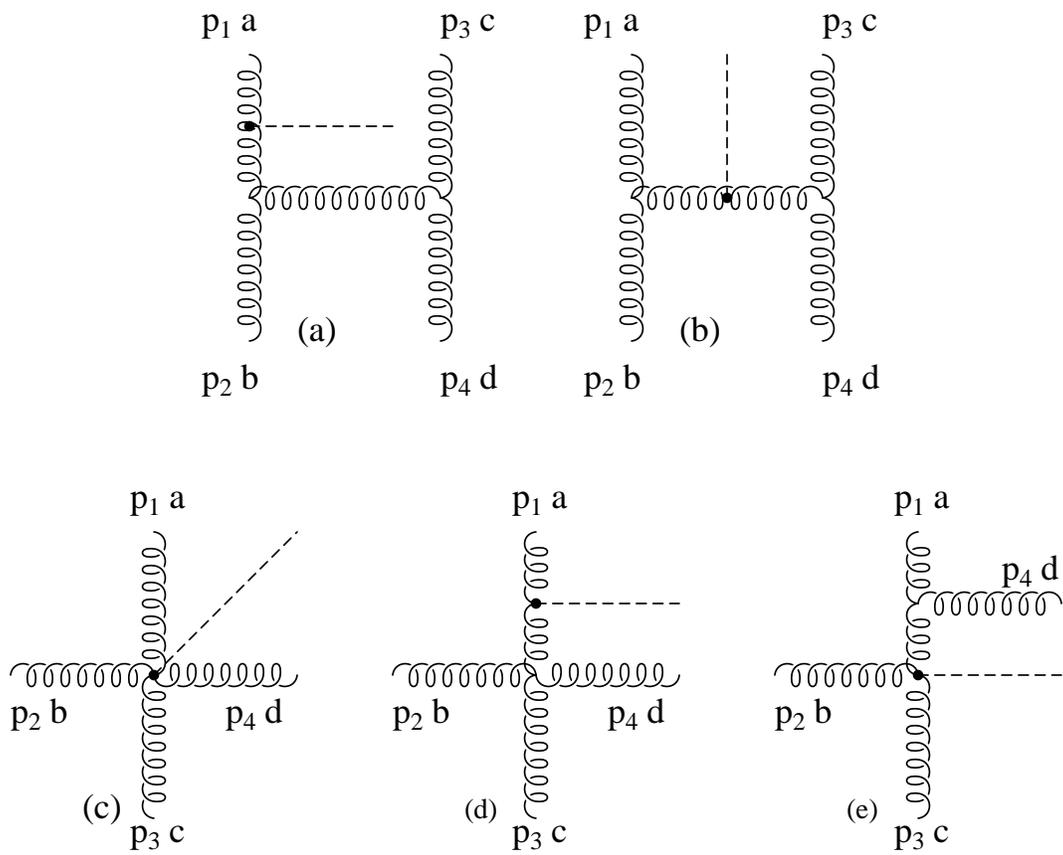
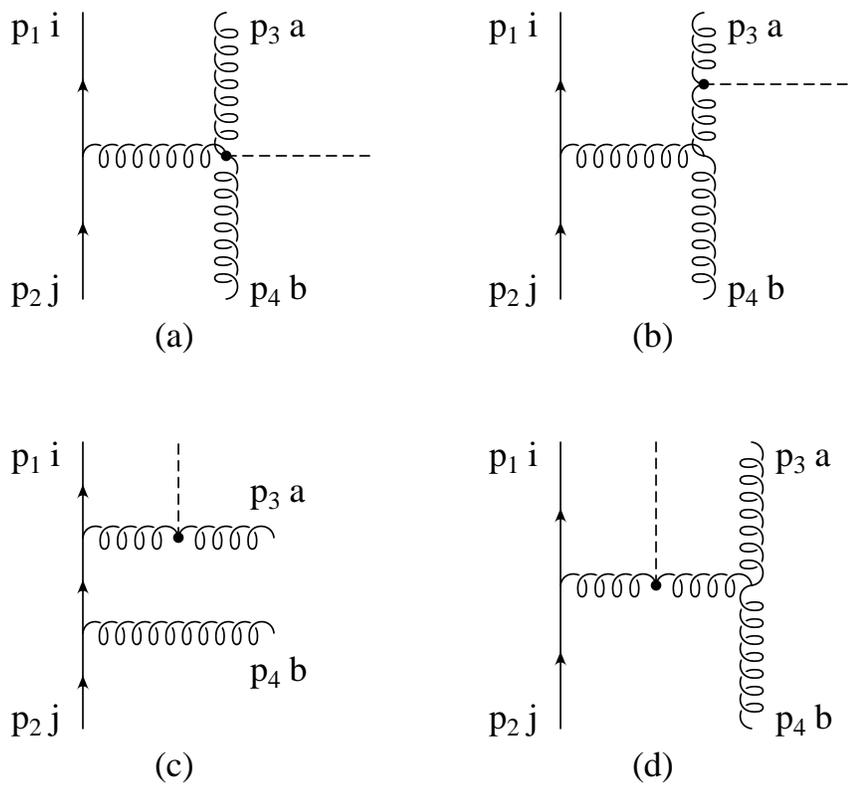
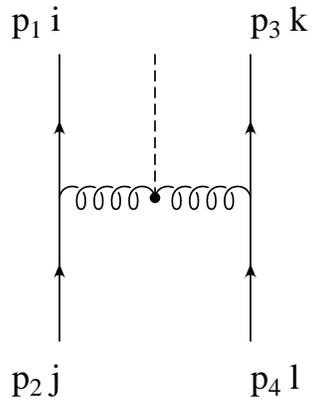
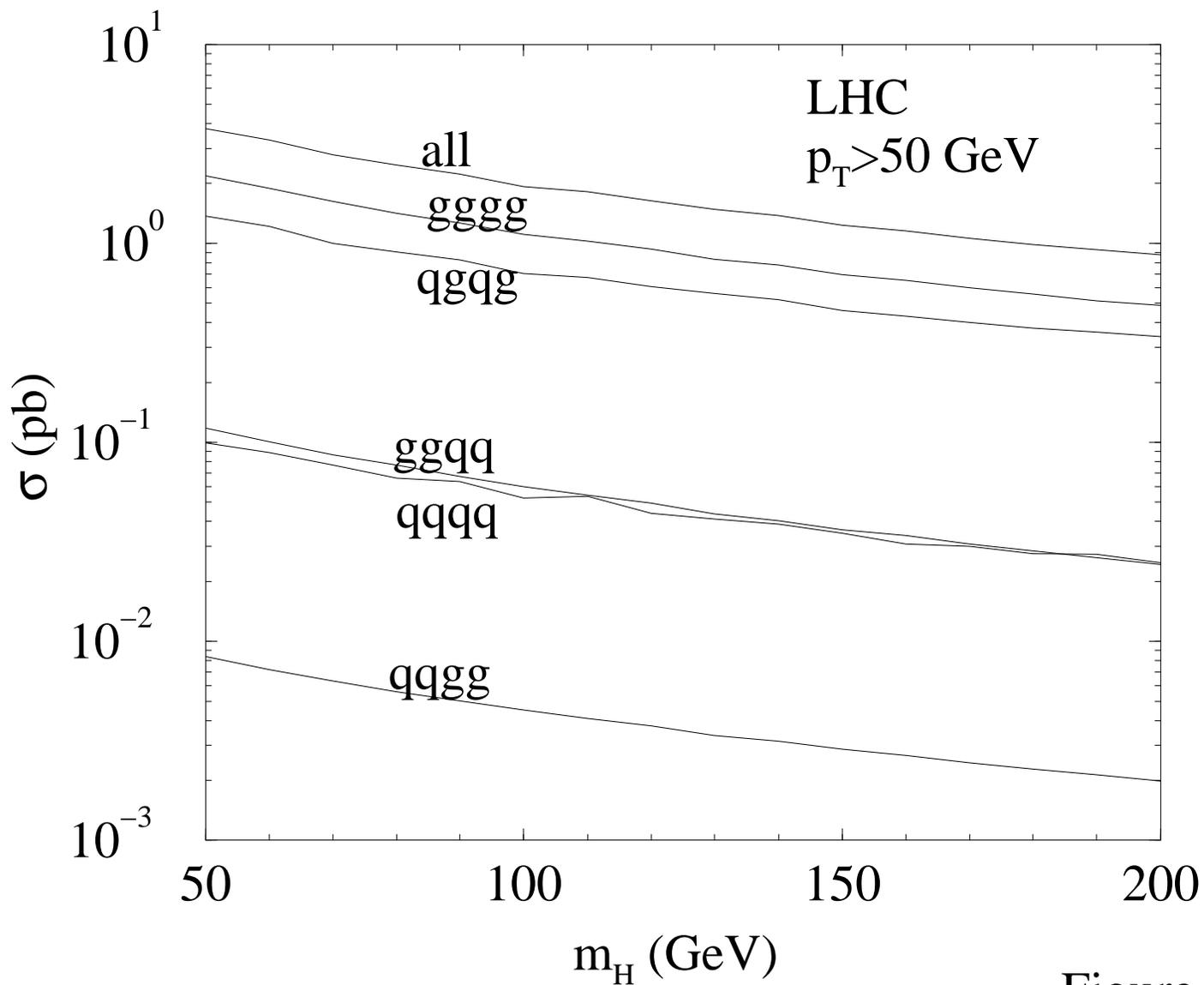
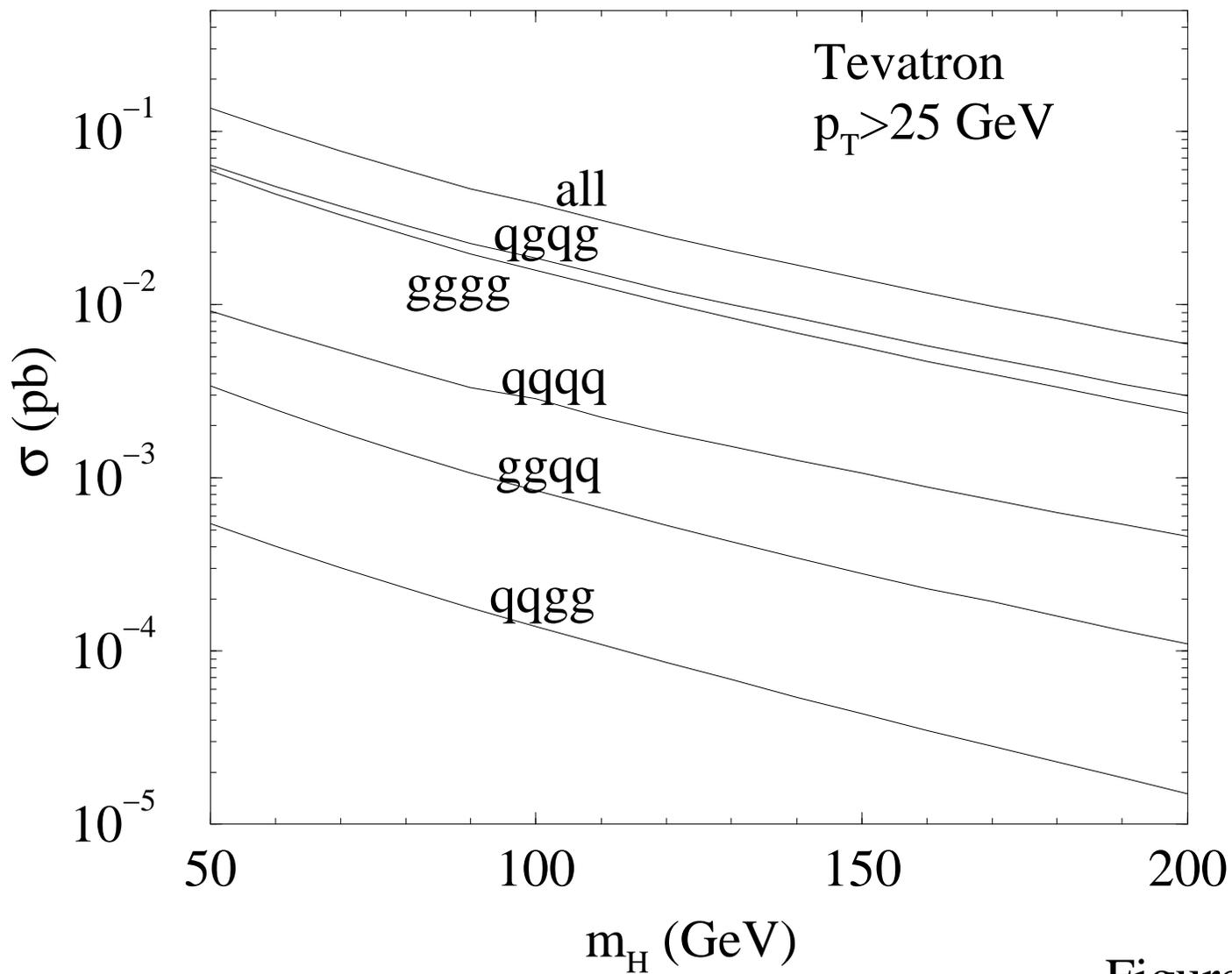
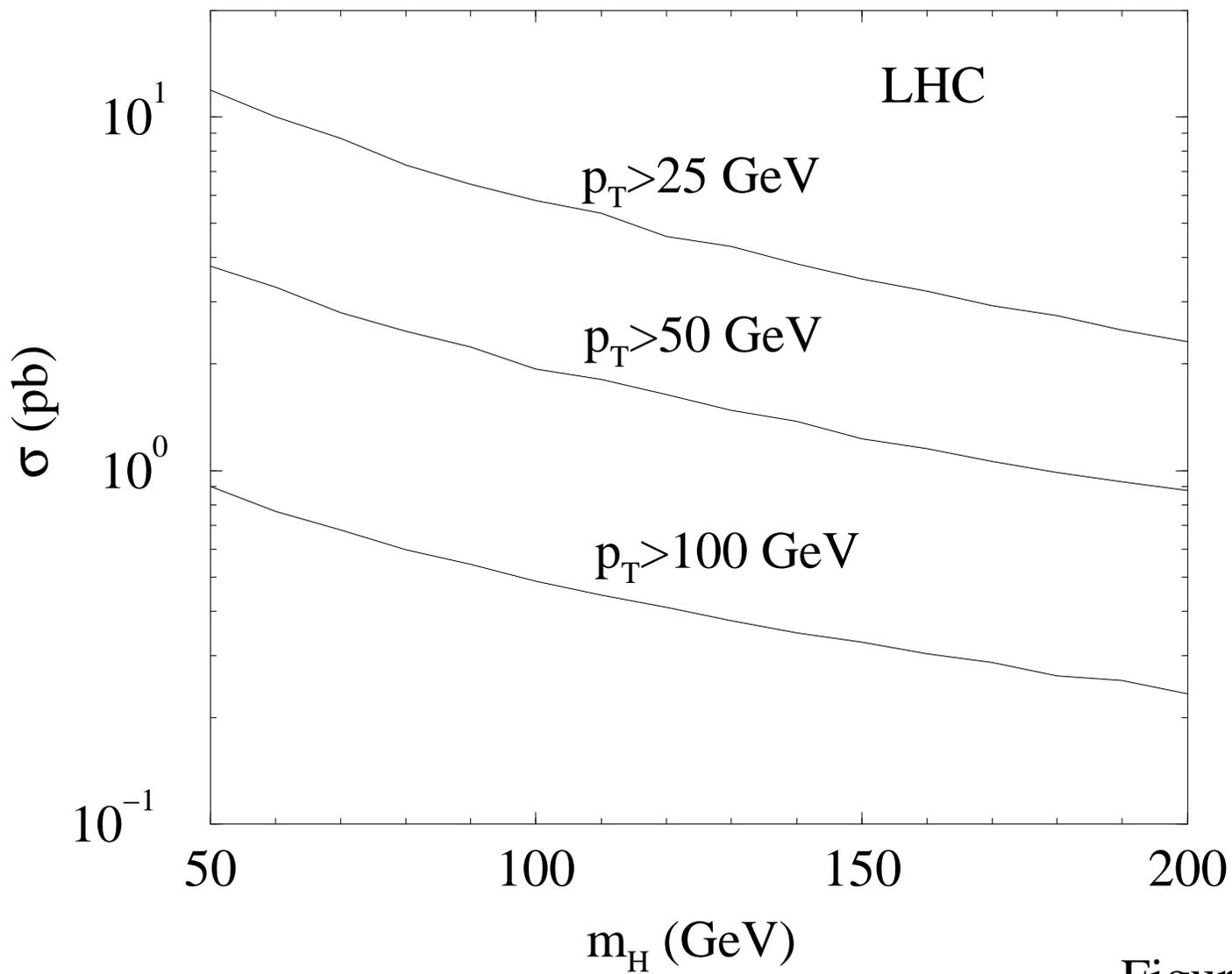
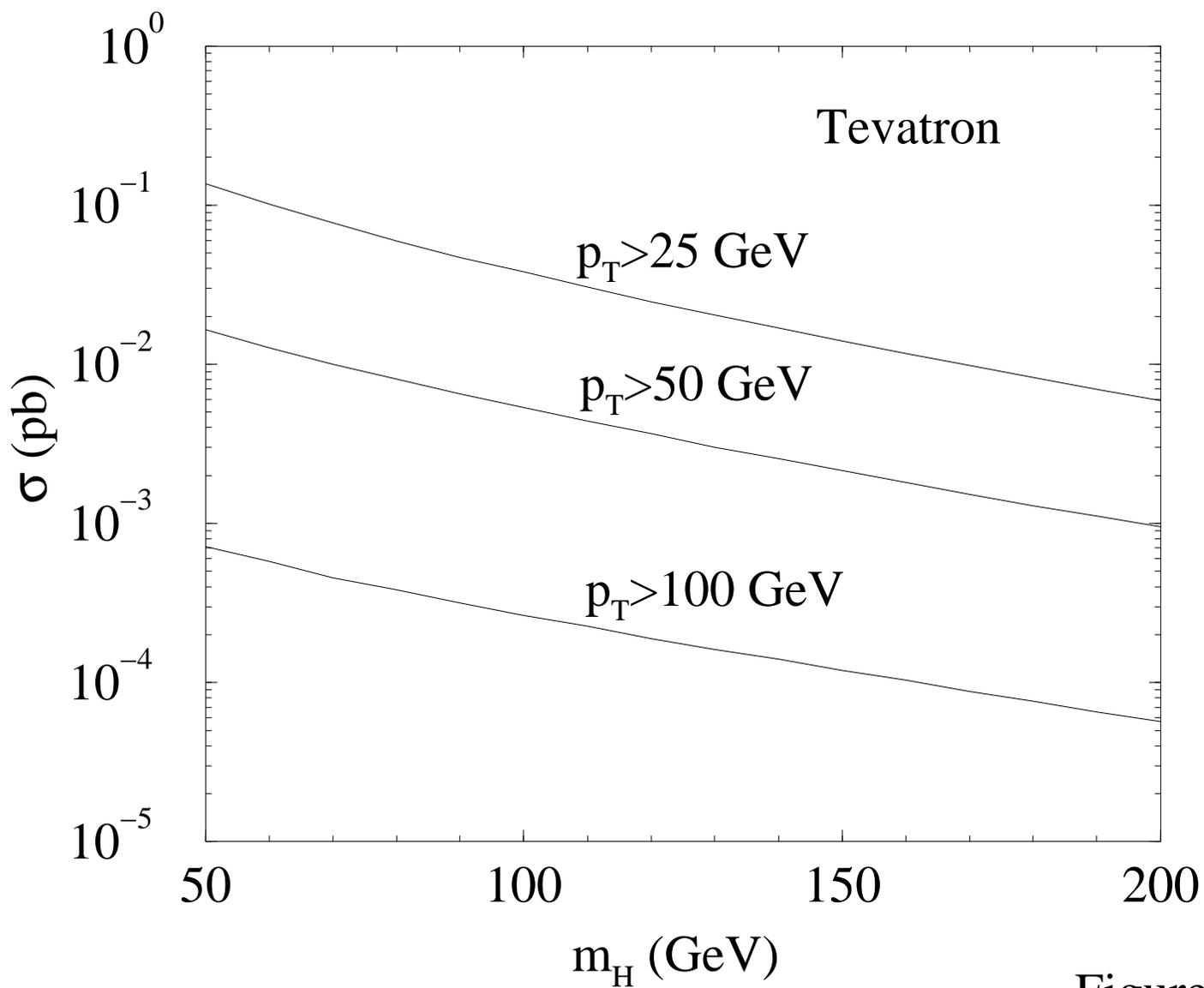
\end{document}